\documentclass{webofc}
\usepackage[varg]{txfonts}   
\usepackage{graphicx}
\usepackage{amsmath}
\usepackage{bm}
\usepackage{hyperref}

\newcommand{\er}{\eqref}

\begin{document}
\title{Chiral medium produced by parallel electric and magnetic fields}
%
%
\subtitle{QCD@work 2016}

\author{\firstname{Marco} \lastname{Ruggieri}\inst{1}\fnsep
\thanks{{\em Speaker.}~~\email{marco.ruggieri@ucas.ac.cn}} \and
        \firstname{Guang Xiong} \lastname{Peng}\inst{1,2}\fnsep\thanks{\email{gxpeng@ucas.ac.cn}} \and
        \firstname{Maxim} \lastname{Chernodub}\inst{3,4}\fnsep\thanks{\email{maxim.chernodub@lmpt.univ-tours.fr}}
}

\institute{College of Physics, University of Chinese Academy of Sciences, 
Yuquanlu 19A, Beijing 100049, China 
\and
           Theoretical Physics Center for Science Facilities, Institute of High Energy Physics, Beijing 100049, China 
\and
           Laboratoire de Math\'ematiques et Physique Th\'eorique UMR 7350, Universit\'e de Tours, 37200 France
 \and
 Laboratory of Physics of Living Matter, Far Eastern Federal University, Sukhanova 8, Vladivostok, Russia
          }

\abstract{%
We compute (pseudo)critical temperature, $T_c$, of chiral symmetry restoration for quark matter in the background
of parallel electric and magnetic fields. This field configuration leads to the production
of a chiral medium on a time scale $\tau$, characterized by a nonvanishing value of the chiral density 
that equilibrates due to microscopic processes in the thermal bath.
We estimate the relaxation time $\tau$ to be about $\approx 0.1-1$ fm/c around the chiral crossover;
then we compute the effect of the fields and of the chiral medium on~$T_c$. 
We find $T_c$ to be lowered by the external fields in the chiral medium.
}
\maketitle
\section{Introduction}
\label{intro}

In recent years it has been realized that in ultrarelativistic heavy ion collisions very strong
electric and magnetic fields can be produced, at least in the very early stages of the 
collision \cite{Skokov:2009qp,Bzdak:2011yy}. In particular the strong magnetic field
and the interplay with chiral (ABJ) anomaly in Quantum Chromodynamics (QCD) ~\cite{Adler:1969gk,Bell:1969ts} 
has suggested the possibility of observation of chiral magnetic effect  (CME) in relativistic
heavy ion collisions~\cite{Kharzeev:2007jp,Fukushima:2008xe}, 
see \cite{Hattori:2016emy} for a recent review and for an exhaustive list of references. 
Naming $n_{R/L}$ number densities of right- and left-handed species,
and ${\cal E}$, ${\cal B}$ a color-electric and color-magnetic field respectively,
CME is related to the generation of chiral density, $n_5=n_R -n_L$ induced by
interaction of quarks with gluon fields characterized by ${\cal E}\cdot{\cal B}\neq 0$ named
the QCD sphalerons.

Because of chiral anomaly it is possible to induce a chiral density also by the interaction of quark matter
with external electric, $E$, and magnetic, $B$, fields with $E\cdot B \neq 0$.
The chiral anomaly will then inject $n_5$ into the system, and this chiral density can equilibrate
thanks to microscopic processes that change quark chirality. The purpose of the study presented here
is to study this equilibration, which leads to the formation of a medium that we name chiral medium,
and to compute its effect on the critical temperature of QCD. The work presented here is mainly based 
on \cite{Ruggieri:2016xww,Ruggieri:2016asg,Ruggieri:2016lrn,Ruggieri:2016ejz,Ruggieri:2016cbq}.

\section{Chiral medium produced by $E$ and $B$}
\label{sec-1}
We consider quark matter in the background of parallel Abelian (electromagnetic) electric, $E$, and magnetic, $B$, 
fields~\cite{Ruggieri:2016xww,Ruggieri:2016asg,Ruggieri:2016lrn}
(see also \cite{Warringa:2012bq,Babansky:1997zh,Cao:2015cka}). In this case
the evolution of $n_5$ with time is given by
\begin{equation}
\frac{dn_5}{dt} =- n_5 \Gamma + N_c\frac{ e^2 E\cdot B}{2\pi^2}\sum_f q_f^2 
e^{-\frac{\pi M_q^2}{|q_f eE|}},
\label{eq:gamma}
\end{equation}
where the first term on the right hand side describes relaxation of chiral density due to
chirality-changing processes in the thermal medium, while the second term comes from the ABJ 
chiral anomaly \cite{Adler:1969gk,Bell:1969ts,Warringa:2012bq}.
$M_q$ denotes the quark mass and $q_f$ the electric charge of the flavor $f$ in units of the proton charge.
The quantity $\Gamma$ in Eq.~\er{eq:gamma} corresponds to the rate of the chirality flips while its inverse defines the 
chiral relaxation time,
\begin{equation}
\tau = 1/\Gamma.
\label{eq:tau}
\end{equation}
The physical interpretation of Eq.~\eqref{eq:gamma} is that in the background of $E\parallel B$ fields the ABJ 
anomaly creates a chiral imbalance $n_5$; the growth of $n_5$ 
competes however with chiral relaxation due to microscopic processes 
that flip the chirality of quarks. 
For time $t\approx\tau$ the value of chiral density exponentially equilibrates at:
\begin{equation}
n_5^{\mathrm{eq}} =N_c \frac{e^2 E\cdot B}{2\pi^2}\tau\sum_f   q_f^2 
e^{-\frac{\pi M_q^2}{|q_f eE|}}.
\label{eq:n5eq}
\end{equation}
The main message encoded in Eq.~\eqref{eq:n5eq} is that the combined effect of the external fields
and of the chirality changing processes in the thermal bath is the formation of chiral medium,
characterized by a finite value of chiral density and to which a chiral chemical potential $\mu_5$
can be associated, namely
\begin{equation}
n_5^{\mathrm{eq}} = -\frac{\partial\Omega}{\partial\mu_5},
\label{eq:ne}
\end{equation}
where $\Omega$ corresponds to the thermodynamic potential.

The knowledge of the relaxation time $\tau$ is crucial since it allows to determine the 
equilibrium value of chiral density, $n_{5}^{\mathrm{eq}}$.
In the context of QCD, $\tau$ been computed in \cite{Ruggieri:2016asg} for temperatures
around $T_c$ within an  NJL model.
The microscopic processes considered in \cite{Ruggieri:2016asg} are 
quark-quark scattering mediated by collective modes with the quantum numbers of pions
and $\sigma-$mesons,
hence we refer to these processes as one-pion or one-$\sigma$ exchange for simplicity. 
While referring the interested reader to \cite{Ruggieri:2016asg} for all the technical details,
we recall here the main quantities that have to be computed to estimate $\tau$. The starting point is 
the Boltzmann collision integral,
\begin{equation}
\frac{d f_R(p)}{dt} = \int d\Pi\frac{(2\pi)^4\delta^4(p+k-p^\prime-k^\prime)}
{2E_p}|{\cal M}|^2 F, 
\label{eq:PS2}
\end{equation}
where $f_{R}$ denotes distribution function for the right-handed quarks 
(a similar equation holds for left-handed quarks) and $d\Pi$ corresponds to 
the  momentum space measure,
\begin{equation}
d\Pi = \frac{d^3k}{(2\pi)^3 2E_k}
 \frac{d^3k^\prime}{(2\pi)^3 2E_k^\prime}
  \frac{d^3p^\prime}{(2\pi)^3 2E_p^\prime};
\end{equation}
the kernel $F$ takes into account the population of the incoming and outgoing particles in the process,
\begin{eqnarray}
F(p,k,p^\prime,k^\prime)&=&f_L(p^\prime)f_L(k^\prime)[1-f_R(p)][1-f_R(k)]
-f_R(p)f_R(k)[1-f_L(p^\prime)][1-f_L(k^\prime)],
\label{eq:op2}
\end{eqnarray}
for the scattering of two incoming $R$ quarks giving two outgoing $L$ quarks,
with $f_{R/L}(p)= (1 + e^{\beta\omega_{\pm}})^{-1}$ and $\omega_s = \sqrt{(p+s\mu_5)^2 + M_q^2}$ with $s=\pm 1$. 
In \cite{Ruggieri:2016asg}  any dependence of $\tau$ on $E$ and $B$ has been neglected for simplicity,
and an expansion to the lowest nontrivial order in $\mu_5/T$ has been used
to further simplify the numerical evaluation of the collision integral.

In Eq.\er{eq:PS2} the squared transition amplitude $|{\cal M}|^2$ 
can be computed once a microscopic process for the chirality change has been chosen,
namely the $\pi-$exchange and $\sigma-$exchange in \cite{Ruggieri:2016asg}.
From the definition of $n_5$ we get the equation governing its time evolution, namely
\begin{equation}
\frac{d n_5}{d t} = N_c N_f\int\frac{d^3 p}{(2\pi)^3}\left(\frac{d f_R}{dt} - \frac{d f_L}{dt} \right).
\label{eq:PS1}
\end{equation}
Then from Eq.\er{eq:gamma} with $E=B=0$ we get the collision rate
$
n_5\Gamma=- dn_5/dt
$;
the relation between $n_5$ and $\mu_5$ has been computed in \cite{Ruggieri:2016asg}
within the NJL model. Once the reaction rate $\Gamma$ is known the relaxation time can be computed by Eq.\er{eq:tau}.

\begin{figure}[t!]
\begin{center}
\includegraphics[width=6.5cm]{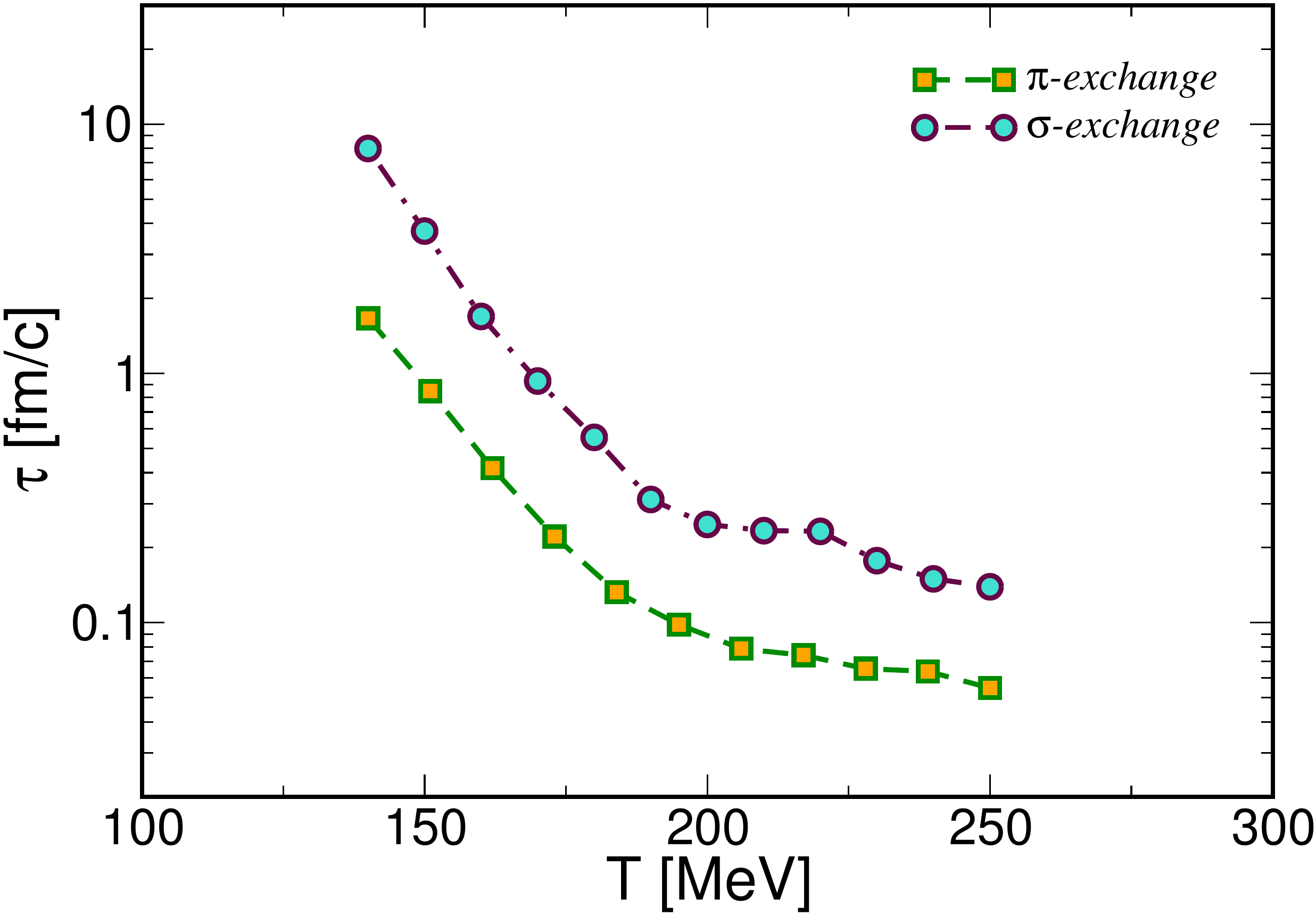}~~~
\includegraphics[width=6.5cm]{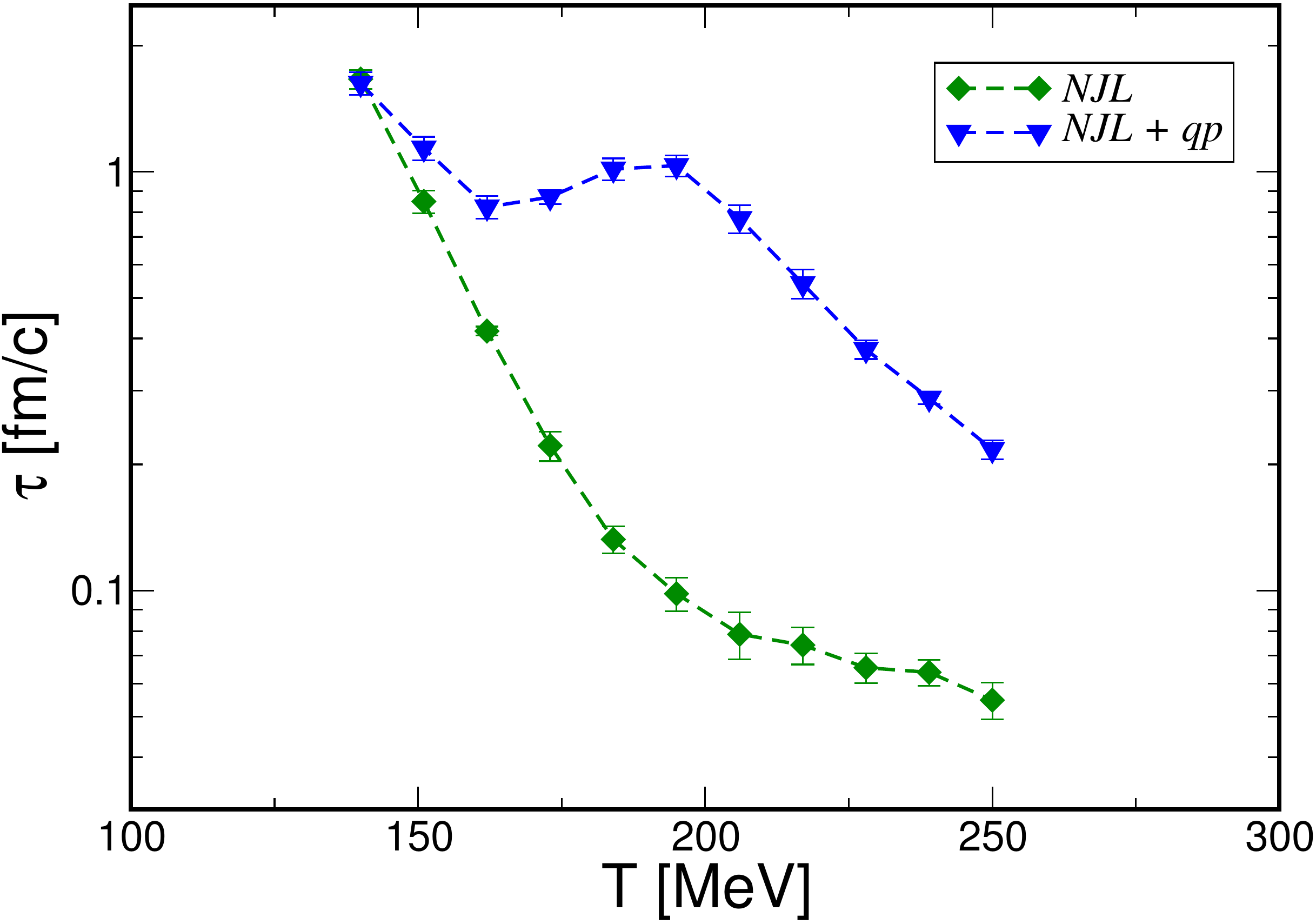}
\end{center}
\caption{\label{Fig:rateCHECK}
Left panel: relaxation time versus temperature in the crossover region for $\pi-$exchange
(squares) and $\sigma-$exchange (circles). Right panel: relaxation time
for $\pi-$exchange for the cases of the pure NJL model (green diamonds) and
of the NJL+quasiparticle model (blue triangles).}
\end{figure}

In the left panel of Fig.~\ref{Fig:rateCHECK} we plot the relaxation time versus temperature in the 
chiral crossover region for $\pi-$exchange (squares) and $\sigma-$exchange (circles).
We notice that the $\sigma-$exchange is less efficient than $\pi-$exchange 
in changing chirality of quarks in the thermal bath, resulting in a larger relaxation time.
This can be understood because of the larger thermal mass of the $\sigma-$meson with respect to pion mass.

An interesting feature of the results shown in Fig.~\ref{Fig:rateCHECK} is that 
the relaxation time decreases with temperature.
This feature might sound counterintuitive: naively a chirality change in the system should be related
to the quark mass operator, that in the NJL model is related to $M_q$ which becomes smaller with temperature, 
implying $\Gamma$ becomes also smaller. 
However, while this naive argument works well for the transition matrix element,
it does not necessarily work for $\Gamma$: in fact the latter is determined also by 
the phase space available for collisions, which instead becomes larger with temperature 
because quark mass becomes smaller and distribution functions broader.
As a result, the collision rate increases with temperature and the relaxation time becomes smaller.
For a more detailed discussion see \cite{Ruggieri:2016asg}.

In \cite{Ruggieri:2016asg} the NJL model calculation of $\tau$ has been compared with a model calculation
where the quark mass function at large temperature is taken to be that of a quasiparticle model \cite{Plumari:2011mk}.
The latter is assumed to take into account many-body effects that are ignored within the NJL model:
in particular, the quasiparticle mass is larger than temperature, at least
for $T\approx T_c$, while the NJL mass becomes much smaller than the temperature itself
because of chiral symmetry restoration. 
The smooth transition from the NJL mass at low temperature to the quasiparticle mass
at large temperature introduced in \cite{Ruggieri:2016asg} models a phase space
volume reduction at $T\approx T_c$, hence potentially lowering $\Gamma$ and increasing $\tau$
at the crossover. 
In the right panel of Fig.~\ref{Fig:rateCHECK} we plot
the relaxation time for $\pi-$exchange computed within the NJL model (green diamonds)
and the NJL+quasiparticle model (blue triangles).
The main difference of the relaxation time computed within the NJL+quasiparticle
model  in \cite{Ruggieri:2016asg}, with respect to the one obtained within the pure NJL model, 
is that in the former case $\tau$ stays almost constant at the crossover rather than
decrease as in the pure NJL model calculation; moreover, 
its order of magnitude is $\tau\approx 1$ fm/c while for the NJL calculation it drops down
to $\tau\approx 0.1$ fm/c for $T>T_c$. Clearly the merging of NJL and quasiparticle models
of  \cite{Ruggieri:2016asg} is too naive: nevertheless the results about $\tau$ show that
the effects of the many body interactions usually neglected in NJL model calculations can have
important effects on the computation of physical quantities around and above $T_c$,
and more serious studies about these effects should be performed.

\section{Critical temperature for chiral symmetry restoration}

\begin{figure}
\begin{center}
\includegraphics[width=6.5cm]{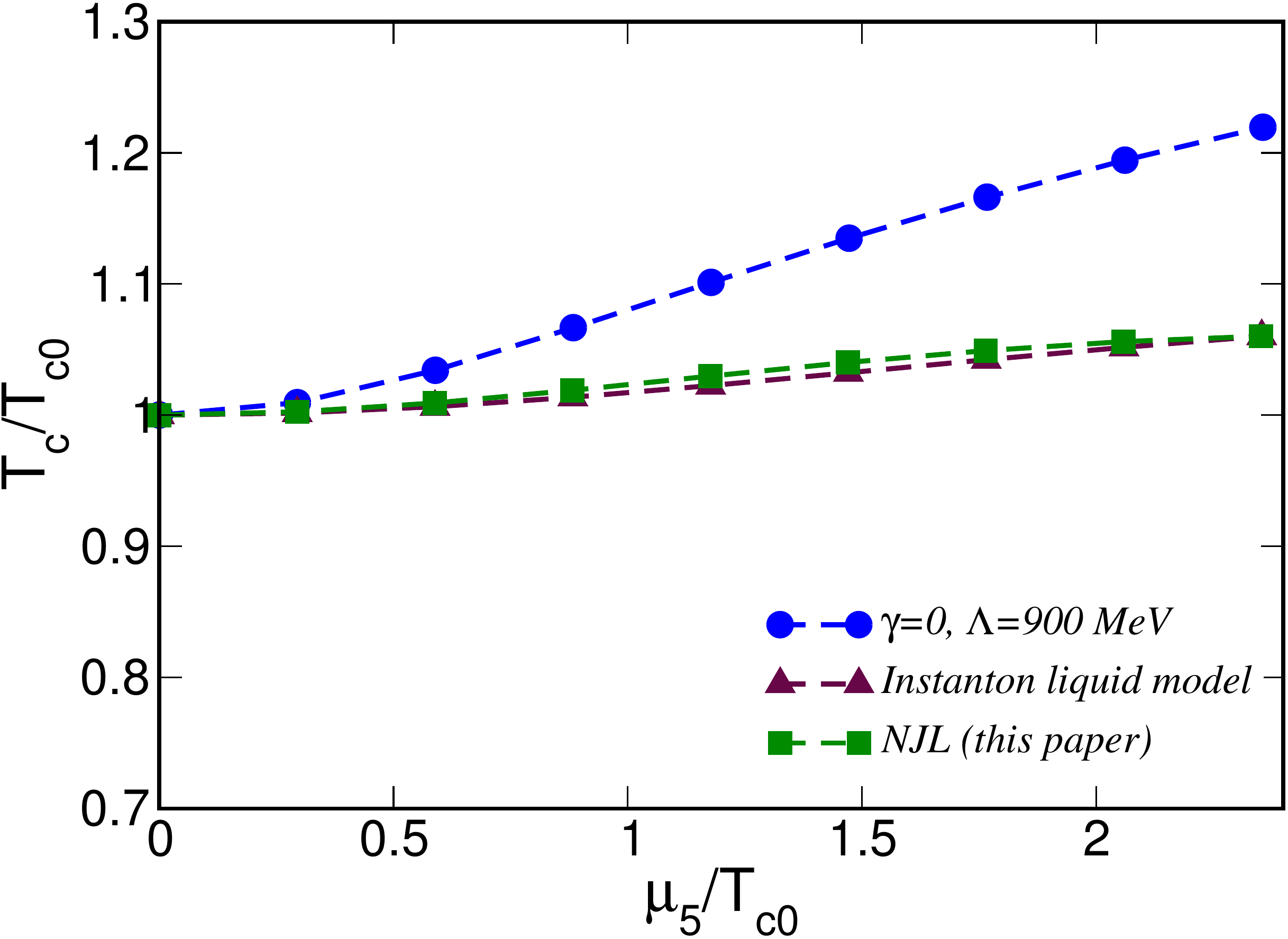}~~~
\includegraphics[width=6.5cm]{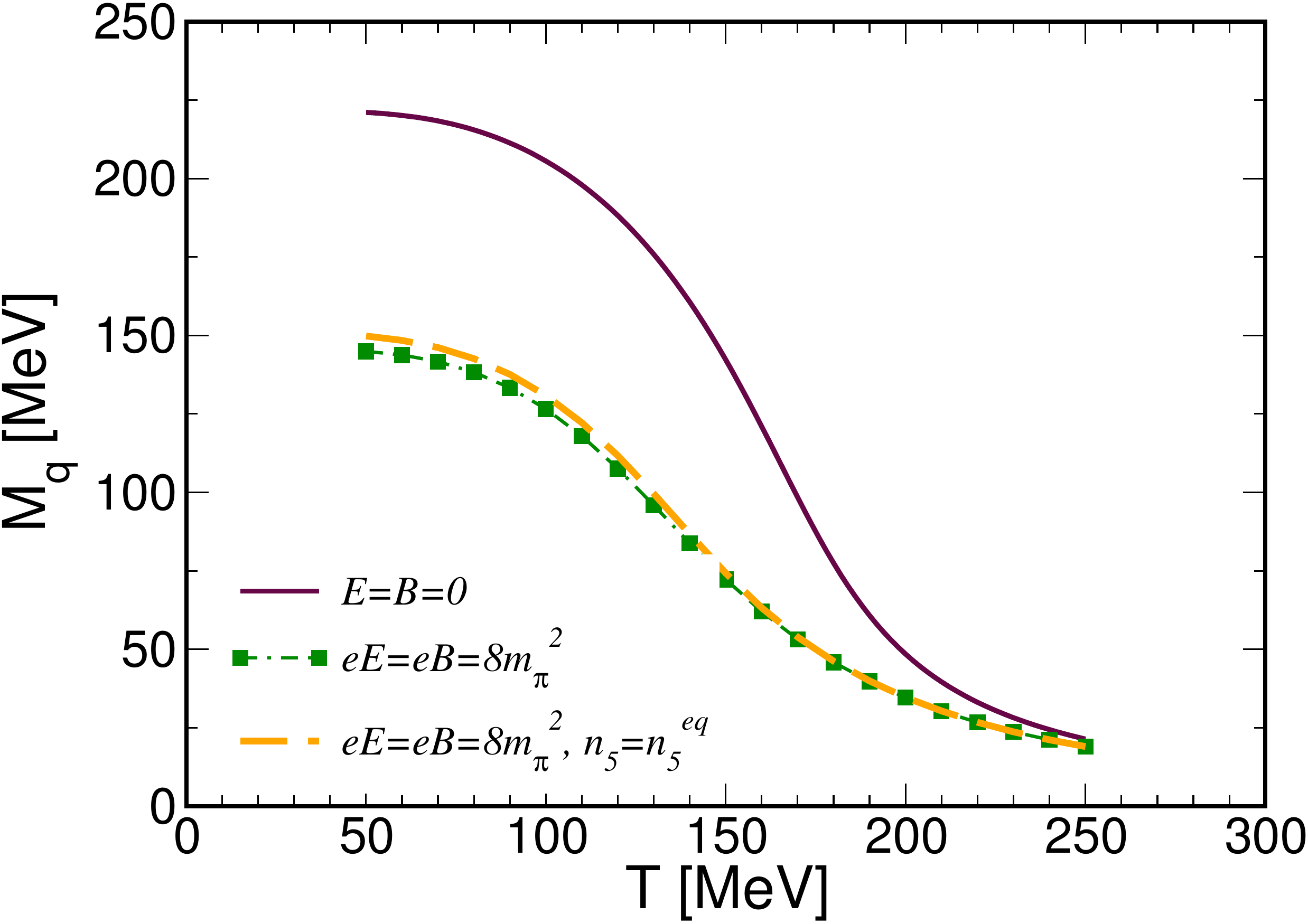}
\end{center}
\caption{\label{Fig:MASS}
Left panel: Critical temperature versus $\mu_5$. Data taken from \cite{Ruggieri:2016ejz}.
Right panel: $M_q$ versus temperature for the case
$eE=eB=8 m_\pi^2$ and $n_5=0$ (green dots), same values of $E$ and $B$ but with
$n_5$ given by its equilibrium value Eq.~\eqref{eq:n5eq} (orange dashed line). For comparison 
we plot by indigo solid line the data corresponding to $E=B=0$. Adapted from \cite{Ruggieri:2016lrn}.}
\end{figure}

A natural question that arises is: can the formation of the chiral medium
lead to any observable effect on equilibrium properties of the bulk? 
In particular, in our study we are interested to the  QCD critical temperature in presence of the electric and 
magnetic fields, therefore we can ask whether the equilibrated $n_5$ can affect $T_c$.
The answer is that in principle this is possible.
To elaborate more on this point
we plot in the left panel of Fig.~\ref{Fig:MASS} the critical temperature
versus $\mu_5$ computed in \cite{Ruggieri:2016ejz} within NJL models with several interaction kernels
for $E=B=0$.
The results in Fig.~\ref{Fig:MASS}
show that $T_c$ increases with $\mu_5$: the chiral chemical potential acts as a catalyzer of
chiral symmetry breaking, see also \cite{Yu:2015hym,Braguta:2015owi,Braguta:2015zta,Braguta:2016aov}. 
Since in the context of our study the chemical potential is produced by the external fields,
the net effect of the field on $T_c$ has to be considered taking into account in the presence the equilibrated $n_5$.

The effect of $\mu_5$ on $T_c$ is opposite to the one induced by the fields.
In fact on the right panel of Fig.~\ref{Fig:MASS} we plot the constituent quark mass versus temperature
for $eE=eB=8m_\pi^2$, for the cases in which we ignore the chiral density at equilibrium
(green squares).
We also plot the data corresponding to $E=B=0$ by the solid indigo line.
Data have been obtained from \cite{Ruggieri:2016lrn}.
The results collected in the right panel of Fig.~\ref{Fig:MASS} show that the fields
lower both $M_q$ and $T_c$ in comparison with the $E=B=0$ case: the fields
thus act as inhibitors of spontaneous chiral symmetry breaking, see also \cite{Babansky:1997zh}. 

Given the different response of $T_c$ to the fields on the one hand,
and to $\mu_5$ on the other hand, it is important to consider both of the effects
to make a firm computation of $T_c$ in presence of $E\parallel B$. This program has been accomplished
in \cite{Ruggieri:2016lrn} where the QCD critical temperature in the background of $E\parallel B$
has been computed taking into account the equilibration of chiral density. In the 
right panel of Fig.~\ref{Fig:MASS} we plot by orange dashed line $M_q$ versus temperature 
for $eE=eB=8m_\pi^2$ for the case in which the equilibrated chiral density has been taken
into account, by solving simultaneously the gap and the number equations within the NJL model.
As expected, the effect of the chiral medium is to increase the value of $M_q$
with respect to the $n_5=0$ case, compare the orange and the green lines.
On the other hand, the net effect of the chiral medium is not so strong, 
since the increase of $M_q$ due to $n_5\neq 0$ is always within the 10$\%$ of the $n_5=0$ result;
as a consequence we conclude the effect of $E\parallel B$ is much stronger than the one
induced by $n_5$ and the net effect of the fields will be a lowering of $T_c$.
Similar results have been recently obtained in \cite{Ruggieri:2016xww}.

In \cite{Ruggieri:2016lrn} the picture drawn here has been completed by the computation of the bulk structure
of the chiral medium at equilibrium: in particular, it has been found that the chiral chemical potentials for $u$
and $d$ quarks are different, a result that can be easily understood because $n_5$ is 
produced by the electromagnetic coupling of quarks to $E$ and $B$ so a dependence on
the quark electric charge is expected. Hence the external fields
induce an imbalanced population of $u$ and $d$ quarks, which might have some effect on the charged pion
condensation.

\begin{figure}
\begin{center}
\includegraphics[width=6.5cm]{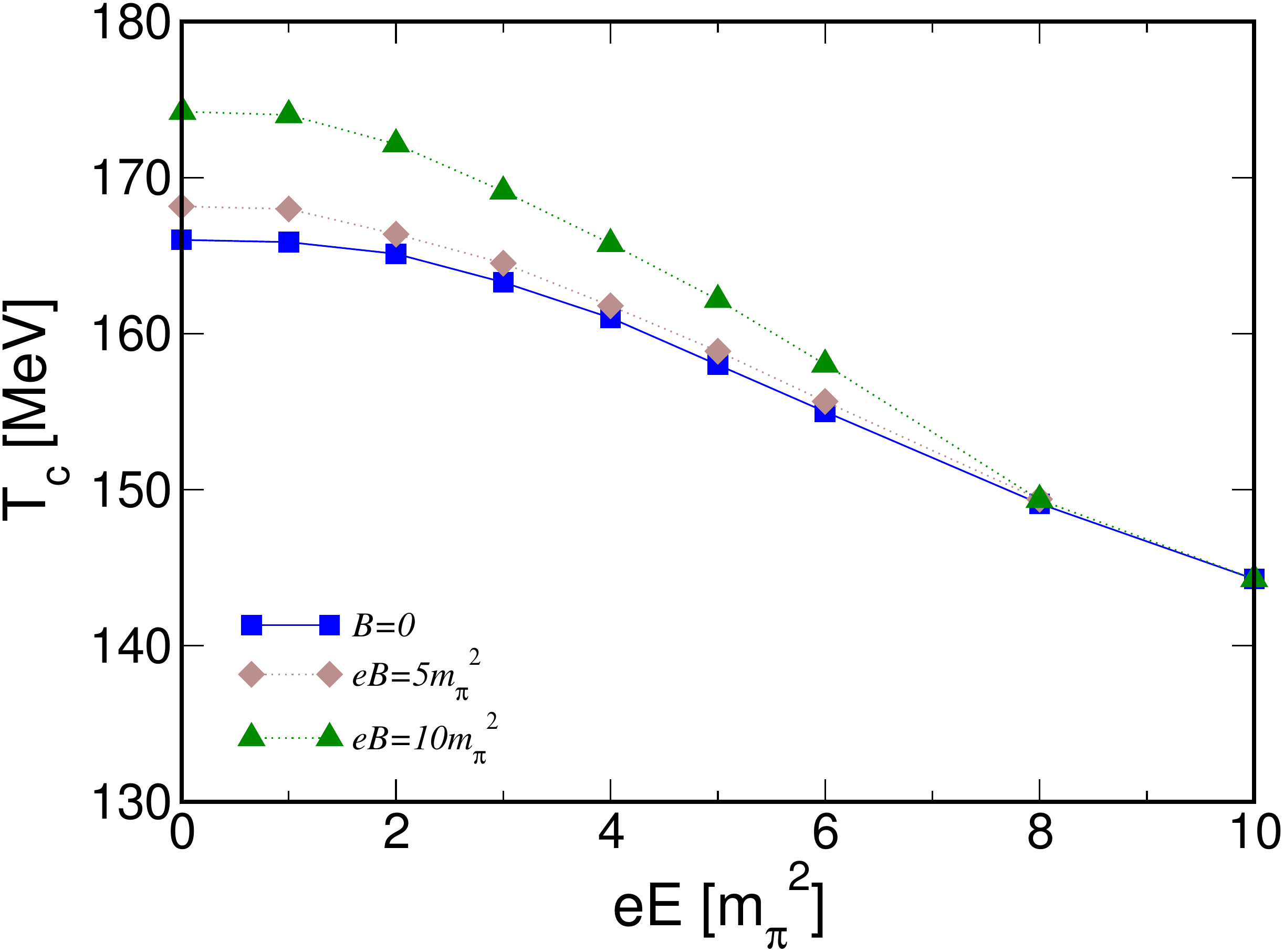}
\end{center}
\caption{\label{Fig:TcVEE}Critical temperature versus electric field strength for several values of $B$.
Adapted from \cite{Ruggieri:2016lrn}.}
\end{figure}
In Fig.~\ref{Fig:TcVEE}, adapted from \cite{Ruggieri:2016lrn}, we plot the critical temperature
as a function of $E$ for several values of $B$. The general trend of the results shown in the figure
is that the electric field lowers $T_c$. An interesting feature of data is that even in the case of the
largest magnetic field considered in the figure, which increases increase $T_c$
with respect to the case $B=0$, a modest electric field $eE\approx 4 m_\pi^2$ is enough to
bring $T_c$ down to the zero field case. Hence the conclusion is that $E\parallel B$ acts as 
an inhibitor of chiral symmetry breaking, as long as $E$ is not much smaller than $B$.

\section{Conclusions}
In this talk we have discussed our results on chiral symmetry restoration at finite temperature,
for quark matter in the presence of parallel electric, $E$, and magnetic, $B$, fields. 
This background leads to the formation of a chiral medium, namely a medium
with a finite chiral density at equilibrium, $n_5=n_R - n_L$, because of the combined effect of the chiral anomaly
which pumps $n_5$ into the system, and chirality changing microscopic processes in the thermal bath
that instead suppress the formation of $n_5$. The characteristic time scale necessary for the formation 
of this medium corresponds approximately to the relaxation time of chiral density, $\tau$.
We have estimated $\tau$ around the chiral crossover of QCD,
finding $\tau\approx 0.1-1$ fm/c. 

Then we have used this value of $\tau$ to
compute $n_5$ at equilibrium simultaneously with the in-medium chiral condensate
at finite temperature, eventually computing the critical temperature, $T_c$,
as a function of the external field magnitude.
Our result is summarized in Fig.~\ref{Fig:TcVEE} where we plot $T_c$ versus the electric field 
strength for several values of the magnetic field; the critical lines have been computed by taking into
account the value of $n_5$ at equilibrium computed self-consistently by virtue of Eq.~\eqref{eq:n5eq}.
We find that the $E\parallel B$ background acts as an inhibitor of chiral symmetry breaking,
at least for values of $E$ not too smaller than $B$.

An interesting follow up of the work presented here might be the inclusion of neutral pion
condensation which is known to be induced by the chiral anomaly \cite{Cao:2015cka};
also the study of charged pion condensation induced by the mismatch of $u$ and $d$
chemical potentials in the chiral medium might be worth to be studied. We plan to report on these subjects in the
near future.

{\em Acknowledgements}. M. R and G. X. P. would like to thank the 
CAS President's International Fellowship Initiative (Grant No. 2015PM008), 
and the NSFC projects (11135011 and 11575190). M. R. acknowledges discussions
with V. Greco, Z. Y. Lu and F. Scardina.

\end{document}